\documentclass[usenatbib]{mn2e}
\usepackage{natbib,psfig,color}
\catcode`\@=11
\def\gsim{\ifmmode{\mathrel{\mathpalette\@versim>}}
    \else{$\mathrel{\mathpalette\@versim>}$}\fi}
\def\lsim{\ifmmode{\mathrel{\mathpalette\@versim<}}
    \else{$\mathrel{\mathpalette\@versim<}$}\fi}
\def\@versim#1#2{\lower 2.9truept \vbox{\baselineskip 0pt \lineskip
    0.5truept \ialign{$\m@th#1\hfil##\hfil$\crcr#2\crcr\sim\crcr}}}
\catcode`\@=12

\newcommand{\beq}{\begin{equation}}
\newcommand{\eeq}{\end{equation}}
\newcommand{\azero}{{a_0}}
\def\Psistar{\Psi_*}

\def\phistar{\phi_*}

\def\rhohalf{\rho_{\rm half}}
\def\rhalf{r_{\rm half}}
\def\NMODY{{\sc n-mody}}
\def\FVFPS{{\sc fvfps}}
\newcommand{\phiN}{\phi^{\rm N}}

\newcommand{\PsistarN}{\Psistar^{\rm N}}
\newcommand{\phistarN}{\phistar^{\rm N}}

\newcommand{\gv}{{\bf g}}
\newcommand{\gvN}{{\bf g}^{\rm N}}
\newcommand{\Sv}{{\bf S}}
\newcommand{\rhostar}{\rho_*}
\newcommand{\varrhostar}{\varrho_*}
\newcommand{\Mstar}{M_*}
\newcommand{\Mstarten}{M_{*,10}}
\newcommand{\tstar}{t_*}
\newcommand{\cafin}{(c/a)_{\rm fin}}

\newcommand{\tdyn}{t_{\rm dyn}}

\newcommand{\rstar}{r_*}
\newcommand{\ra}{r_{\rm a}}

\newcommand{\vstar}{v_*}

\newcommand{\Npart}{N_{\rm part}}

\newcommand{\xis}{\xi_{\rm s}}
\newcommand{\xihalf}{\xi_{\rm half}}
\newcommand{\xihalfs}{\xi_{\rm half,s}}
\newcommand{\Tr}{T_{\rm r}}
\newcommand{\Tt}{T_{\rm t}}

\newcommand{\vcMOND}{v_{\rm c}^{\rm M}}
\newcommand{\vcN}{v_{\rm c}^{\rm N}}

\def\ras{r_{\rm as}}
\def\rac{r_{\rm ac}}

\def\d{\rm d}

\def\en{{\mathcal{E}}}

\newcommand{\bey}{\begin{eqnarray}}
\newcommand{\eey}{\end{eqnarray}}
\newcommand{\Myr}{\, {\rm Myr} }

\newcommand{\kpc}{\, {\rm kpc} }
\newcommand{\Msun}{M_\odot}

\newcommand{\kms}{\, {\rm km \, s}^{-1} }

\newcommand{\xv}{{\bf x}}

\newcommand{\fN}{f_{\rm N}}

\newcommand{\fMOND}{f_{\rm M}}

\def\rt{r_{\rm t}}

\def\xv{{\bf x}}

\def\xv{{\bf x}}

\def\Ttheta{T_{\vartheta}}
\def\Tphi{T_{\varphi}}

   \title[Radial-orbit instability in MOND]
         {Radial-orbit instability in modified Newtonian dynamics}
\author[]{}

   \author[Nipoti et al.]
          {Carlo Nipoti,$^1$\thanks{E-mail: carlo.nipoti@unibo.it}
           Luca Ciotti$^1$ and Pasquale Londrillo$^2$
           \\ $^1$Astronomy Department, University of Bologna, 
                       via Ranzani 1, I-40127 Bologna, Italy
           \\ $^2$INAF-Bologna Astronomical Observatory, 
                       via Ranzani 1, I-40127 Bologna, Italy}

\date{Accepted 2011 March 2.  Received 2011 February 28; in original form 2011 January 31}
\pubyear{2011}

\begin{document} 
\maketitle

\begin{abstract}
  The stability of radially anisotropic spherical stellar systems in
  modified Newtonian dynamics (MOND) is explored by means of numerical
  simulations performed with the $N$-body code \NMODY.  We find that
  Osipkov-Merritt MOND models require for stability larger 
    minimum anisotropy radius than equivalent Newtonian systems (ENSs)
    with dark matter, and also than purely baryonic Newtonian models
    with the same density profile. The maximum value for stability of
    the Fridman-Polyachenko-Shukhman parameter in MOND models is lower
    than in ENSs, but higher than in Newtonian models with no dark
  matter.  We conclude that MOND systems are substantially more prone
  to radial-orbit instability than ENSs with dark matter, while they
  are able to support a larger amount of kinetic energy stored in
  radial orbits than purely baryonic Newtonian systems. An explanation
  of these results is attempted, and their relevance to the MOND
  interpretation of the observed kinematics of globular clusters,
  dwarf spheroidal and elliptical galaxies is briefly discussed.
\end{abstract}


\section{Introduction}
\label{secint}

Modified Newtonian dynamics (MOND) was originally proposed by
\citet{Mil83} to explain the rotation curves of disk galaxies without
invoking the presence of dark matter (DM) and, over the years, it has
been successful at reproducing the observed kinematics of several
galaxies \citep[e.g.][and references therein]{San02,San07,Swa10}.
However, to some extent, the MOND and DM interpretations of the
kinematics of galaxies can be degenerate. For instance, a MOND
rotation curve can be also described in the Newtonian gravity invoking
a DM halo such that the total gravitational field is the same as the
MOND one.  More generally, given a MOND system, it is possible to
  construct the {\it equivalent Newtonian system} (ENS), i.e. the
Newtonian system with DM in which the visible matter has the same
phase-space distribution as in the MOND system
\citep*{Mil86,Mil01,Nip07b,Nip07c,Nip08}. It should be noted, however,
that the physical viability of the ENS is not guaranteed, as for some
configurations the density of the associate DM halo turns out to be
negative \citep{Mil86}.  Though a MOND system and its ENS are, by
construction, indistinguishable from a {\it kinematic} point of view
(for instance, as far as the rotation curve\footnote{The MOND rotation
  curve of a disk galaxy can be always reproduced with a {\it
    spherical} DM halo and Newtonian gravity. However, the vertical
  kinematics differs in the two cases \citep{Nip07b}, because the DM
  halo of the ENS of a disk galaxy is non-spherical.} or the
velocity-dispersion profile are concerned), in general they are not
identical from a {\it dynamical} point of view: for instance,
two-body relaxation, dynamical friction and galaxy merging act
differently in the two cases \citep{CioB04,Nip07c,Nip08}.

As already recognized, MOND and ENSs might differ substantially also
in terms of stability.  As one of the original motivations for
invoking DM halos in disk galaxies was that purely baryonic Newtonian
disks are prone to bar-like instability \citep{Ost73}, it is not
surprising that the study of dynamical stability in MOND has focused
mainly on disks \citep{Mil89,Chr91,Bra99}.  Here we consider instead
the so-called radial-orbit instability, which is relevant to
pressure-supported stellar systems.  As well known, in the context of
Newtonian gravity the amount of radial orbits in stellar systems is
limited not only by the requirement of phase-space consistency
\citep[i.e. positivity of the distribution function; see,
  e.g.,][]{CioP92,AnE06,CioM09,CioM10a,CioM10b} but also by the fact
that very radially anisotropic spherical systems are unstable
\citep{Hen73,Pol81,Fri84,MerA85,Bar86,May86,Pal87,Dej88,Sah91,Wei91,Ber94,Hjo94,Mez97,Tre06,Bar09}.
In Newtonian gravity DM halos tend to have a stabilizing effect
against radial-orbit instability \citep*{Sti91,Mez98,Nip02}, which
suggests that MOND systems might be more prone to this kind of
instability than their ENSs.  If confirmed, this could provide a
discriminant between the two theories when interpreting the velocity
dispersion profiles in pressure-supported stellar systems. For these
reasons, in this work we use $N$-body simulations to explore the
stability of radially anisotropic MOND spherical systems and, for
comparison, of their ENSs and of purely baryonic Newtonian systems
with the same density distributions and anisotropy profiles.

The paper is organized as follows. In Section~\ref{secmod} the galaxy
models used in the simulations are described. Their phase-space
consistency is discussed in Section~\ref{seccon} and their stability
in Section~\ref{secsta}.  Section~\ref{secfin} concludes.

\begin{table*}
 \flushleft{
  \caption{Main properties of  the studied families of models.}
\begin{tabular}{llccccc}
Family & Gravity & $(M_{\rm DM}/\Mstar)_{\rm half}$ &
$\rac/\rhalf$ & $\ras/\rhalf$ & $\xis$ &$\xihalfs$ \\
   ~~(1) & ~~~(2) & (3) & (4) & (5) &  (6) & (7) \\ 
\hline
N$_0$                      & Newton   & 0    & 0.092  & $\sim$0.8  & $\sim$1.75 & $\sim$1.34\\
M$_{0}\kappa_{10}$   & MOND    & 0    & 0.092 & $\sim$1.0 & $\sim$2.33 & $\sim$1.26 \\
E$_{0}\kappa_{10}$   & Newton   & 0.76 & 0.092 & $\sim$0.5 & $\sim$3.48 & $\sim$1.83 \\
M$_{0}\kappa_{0.01}$& MOND    & 0    & 0.092 & $\sim$1.0 & $\sim$2.48  & $\sim$1.28\\
E$_{0}\kappa_{0.01}$ & Newton  & 50.1 & 0.092 & $<$0.2 &$>$7.48 & $>$4.06 \\
\hline
N$_1$                     & Newton   & 0    & 0.053 & $\sim$0.8 & $\sim$1.64& $\sim$1.28 \\
M$_{1}\kappa_{10}$ & MOND     & 0      & 0.053 & $\sim$0.8 & $\sim$2.32& $\sim$1.31 \\
E$_{1}\kappa_{10}$  & Newton   & 0.31   & 0.053 & $\sim$0.6 & $\sim$2.70 & $\sim$1.50\\
M$_{1}\kappa_{0.01}$& MOND   & 0    & 0.053 & $\sim$0.9 &$\sim$2.61 & $\sim$1.31\\ 
E$_{1}\kappa_{0.01}$ & Newton & 32.7 & 0.053 & $<$0.2&$>$6.86  & $>$3.64 \\
\hline
\end{tabular}
} \medskip \flushleft{(1) Name of the family of models: N are the
  Newtonian purely baryonic models, M the MOND models, and E the ENS
  models. The first subscript is the value of the parameter $\gamma$
  identifying the stellar density profile (equation~\ref{eqrhostar});
  the second subscript is the value of the internal acceleration ratio
  $\kappa\equiv G\Mstar/\azero\rstar^2$. (2) Gravity law. (3)
  DM-to-stellar mass ratio within the stellar half-mass radius
  $\rhalf$ ($\rhalf\simeq 3.84\rstar$ for $\gamma=0$ and $\rhalf\simeq
  2.41\rstar$ for $\gamma=1$). (4) Normalized critical anisotropy
  radius for consistency, as determined from the GDSAI. Models with
  $\ra <\rac$ are necessarily inconsistent. (5) Normalized minimum
  anisotropy radius for stability.  (6) Maximum value for stability of
  the Fridman-Polyachenko-Shukhman parameter. (7) Maximum value for
  stability of the radial-to-tangential kinetic energy ratio measured
  within $\rhalf$.  In the columns (5-7), upper limits on $\ra$ and lower
  limits on $\xis$ and $\xihalfs$ mean that no instability has been
  detected in the most anisotropic case explored for that family. }
\end{table*}

\section{Galaxy models}
\label{secmod}

We consider MOND as modified gravity in the non-relativistic
formulation\footnote{Recently an alternative non-relativistic
  formulation of MOND, dubbed QUMOND, has been proposed
  \citep{Mil10}.} of \citet{BekM84}, in which the Poisson equation
$\nabla^2\phiN=4\pi G\rho$ is replaced by
\begin{equation}
\nabla\cdot\left[\mu\left({\Vert\nabla\phi\Vert\over\azero}\right)
\nabla\phi\right] = 4\pi G \rho,
\label{eqMOND}
\end{equation} 
where $\Vert ...\Vert$ is the standard Euclidean norm, $\phi$ is
  the MOND gravitational potential, $\azero \simeq 1.2 \times 10^{-10}
  {\rm m}\,{\rm s}^{-2}$ is a characteristic acceleration, and $\mu$
  is the so-called interpolating function \citep[in the present work
    we adopt $\mu(y)=y/\sqrt{1+y^2}$;][]{Mil83}. The MOND gravitational
  acceleration is $\gv=-\nabla\phi$, just as the Newtonian
  acceleration is $\gvN=-\nabla\phiN$.  For a system of finite mass,
  $\nabla\phi\to 0$ as $\Vert\xv\Vert\to\infty$, where $\xv$ is
    the position vector.

As well known, from the Poisson equation and equation~(\ref{eqMOND})
it follows that the MOND and Newtonian gravitational accelerations are
related by ${\mu}(g/\azero) \, \gv = \gvN +\Sv$, where
$g\equiv\Vert\gv\Vert$, and $\Sv$ is a solenoidal field dependent on
the specific $\rho$ considered. Since in general $\Sv\ne0$, and its
expression is unknown {\it a priori}, standard Poisson solvers cannot
be used to develop MOND $N$-body codes, in which
equation~(\ref{eqMOND}) must be solved at each time step
\citep*[see][]{Bra99,Cio06,TirC07}.  In the present work we use our
original MOND $N$-body code \NMODY\ \citep*[][see
  Section~\ref{secsim}]{Nip07a,Lon09}.

The initial conditions of the simulations are $N$-body realizations of
galaxy models with the stellar component described by a spherical
$\gamma$-model \citep{Deh93,Tre94} with density distribution
\begin{equation}
\rhostar (r)= {3-\gamma\over 4\pi}{\Mstar\rstar \over 
r^{\gamma} (\rstar+r)^{4-\gamma}},
\label{eqrhostar}
\end{equation}
where $\Mstar$ is the total stellar mass, $\rstar$ is the scale
radius, and $\gamma$ is the negative of the inner logarithmic density
slope ($0\leq \gamma <3$).  For simplicity we restrict to the cases
$\gamma=0$ and $\gamma=1$ \citep{Her90}; we recall that for
$\gamma\neq 2$ the Newtonian potential is
\begin{equation}
\phistarN(r)={G\Mstar \over \rstar (2-\gamma)}
            {\left[\left({r\over \rstar+r}\right)^{2-\gamma} -1\right]}.
\end{equation}
For these models the MOND potential, required to distribute the
particles in the velocity space, is easily calculated from the
Newtonian one, as in spherical symmetry $\Sv=0$. The ENS associated
with a model of stellar density $\rhostar$ and MOND potential $\phi$
has a {\it total} density $\nabla^2\phi(r) /4\pi G$.  However, the
resulting DM halo density (obtained after subtraction of $\rhostar$)
in principle may be negative or have a non-monotonic radial trend.
Fortunately, it can be proved that the DM halos of the ENS derived
from the spherical $\gamma=0$ and $\gamma=1$ models are everywhere
positive. Instead, possible non-monotonicity of the DM density
distribution makes the consistency of the halo a non-trivial request,
as we will discuss in Section~2.1 \citep*[see also][]{CioMdZ09}.

In order to impose a tunable amount of radial anisotropy on the
initial conditions, we adopt the widely used Osipkov-Merritt (OM)
parametrization \citep[][]{Osi79,Mer85}, in which the stellar
distribution function depends on energy and angular momentum per unit
mass through the variable $Q\equiv \en-{J^2/2\ra^2}$, where $\en
=\Psi-v^2/2$ is the relative energy, $v=||{\bf v}||$ is the velocity
modulus, $\Psi=-\phi$ is the relative potential, $J$ is the angular
momentum modulus per unit mass, and $\ra$ is the so-called anisotropy
radius. The anisotropy parameter \citep[e.g.][]{BT08} is
$\beta(r)=r^2/(\ra^2+r^2)$: for $r \gg\ra$ the velocity dispersion
tensor is radially anisotropic, while for $r \ll \ra $ the tensor is
nearly isotropic.  In the limit $\ra \to\infty$, $Q=\en$ and the
velocity dispersion tensor becomes globally isotropic.

In the purely baryonic Newtonian models, the distribution function 
can be written as
\begin{equation}
\fN(Q)={1\over \sqrt{8}\pi^2}
{\d\over \d Q}\int_0^{Q}{\d\varrhostar \over \d\PsistarN}
{\d\PsistarN\over\sqrt{Q-\PsistarN}},
\label{eq:dfn}
\end{equation}
where $\PsistarN=-\phistarN$ is the Newtonian relative potential and
\begin{equation}
\varrhostar (r)=\left(1+\frac{r^2}{\ra^2}\right)\rhostar (r)
\label{eq:varrhostar}
\end{equation}
is the so-called augmented density \citep[see, e.g.,][]{BT08}.

In the MOND (and in the ENS) cases a similar formula holds, i.e.
\begin{equation}
\fMOND(Q)={1\over \sqrt{8}\pi^2}{\d \over \d Q}\int_{-\infty}^{Q}
{\d\varrhostar\over \d\Psi}{\d\Psi\over\sqrt{Q-\Psi}},
\label{eq:dfm}
\end{equation}
where now $\Psi$ is the MOND relative potential that, by construction,
is also the total potential of the ENS. Note that, at variance with
equation (4), the lower integration limit is now $-\infty$: due to the
far-field logarithmic behaviour of the MOND potential in isolated
systems of finite mass, the distribution function must be positive for
all values of $Q<0$ in analogy with finite-mass Newtonian system in a
total isothermal potential, where stars of all energies are bound
\citep[e.g.][]{CioMdZ09}. Apart from a few exceptions (see
  Section~\ref{seclive}), in the simulations of the ENSs the halo is
  maintained ``frozen'' (i.e. it acts as a fixed external potential),
  so its distribution function is not needed.

\subsection{Consistency}
\label{seccon}

As well known, in (single- or multi-component) OM models it is
possible to determine a critical value of the anisotropy radius of
each density component, $\rac$, so that for $\ra <\rac$ the models are
inconsistent, i.e., $f(Q)<0$ for some admissible value of $Q$. A
necessary condition for consistency of OM models is $d\varrhostar(r)/d
r\leq 0$ at all radii \citep{CioP92,Cio96,Cio99,CioMdZ09}. As recently
proved in Ciotti \& Morganti (2010a,b; see also Van Hese, Baes \&
Dejonghe 2011; An 2011) this condition is rigorously equivalent to the
Global Density Slope-Anisotropy Inequality (GDSAI), i.e. to the
requirement that at each radius the negative of the logarithmic slope
of the stellar density distribution cannot be less than twice the
local value of the anisotropy parameter, $\gamma_*(r)\geq 2\beta (r)$.
Remarkably, it is easy to show that the GDSAI {\it is independent of
  the gravity law holding the system together}, so that it holds not
only for Newtonian multi-component systems (such as the ENS), but also
in the MOND case\footnote{This however is not true for the {\it
    sufficient conditions} for consistency
  \citep[see][]{CioM10a,CioM10b}}.

We applied the GDSAI to our families of models, determining for each
stellar density profile the value of $\rac$ (see Table 1), and the
obtained limits coincide with those determined in \citet{Cio99}. As
these are just necessary conditions for consistency, the positivity of
the distribution function must still be checked numerically for
$\ra\geq\rac$.  We also note that from the inequality
$d\varrhostar(r)/d r\leq 0$ it follows that in case of isotropy
($\ra=\infty$) a spherical density distribution must be necessarily
monotonically decreasing for increasing $r$, and this imposes quite
strong constraints on the DM halos of physically admissible ENS.  If
the halo has a central ``hole'' \citep[which is not unusual for ENSs;
  see][]{Nip07c}, it can be physically realized only if its orbital
distribution is tangentially biased, at least in the internal regions.
In practice, this is not an issue in simulations with frozen DM halos,
but it is an important problem when setting up initial conditions for
simulations in which the DM halo is ``live'' (i.e. modelled with
particles; see Section~\ref{seclive}).

\begin{figure*}
\centerline{ \psfig{file=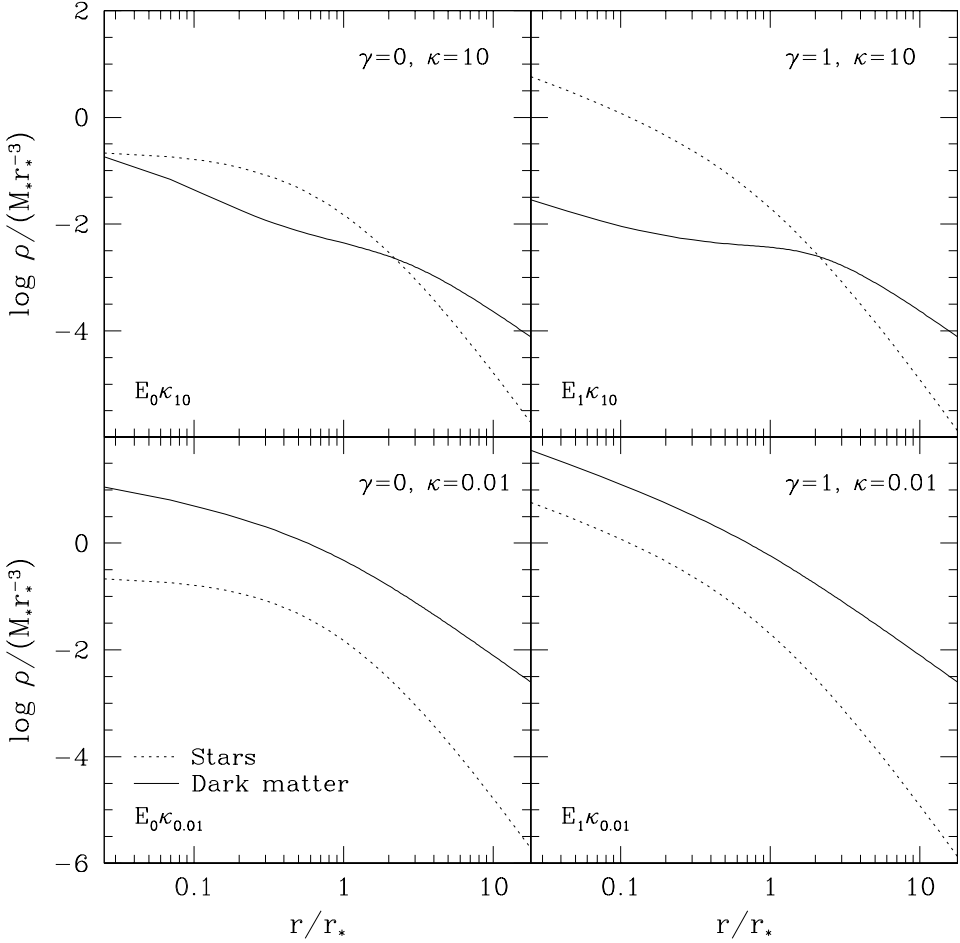,width=0.75\hsize}}
\caption{Stellar (dotted curves) and DM (solid curves) density
  profiles of ENSs corresponding to spherical MOND $\gamma=0$ (left
  column) and $\gamma=1$ (right column) models, for two different
  values of the dimensionless acceleration ratio $\kappa
  =G\Mstar/\azero\rstar^2$.}
\label{figden}
\end{figure*}

\begin{figure*}
\centerline{ \psfig{file=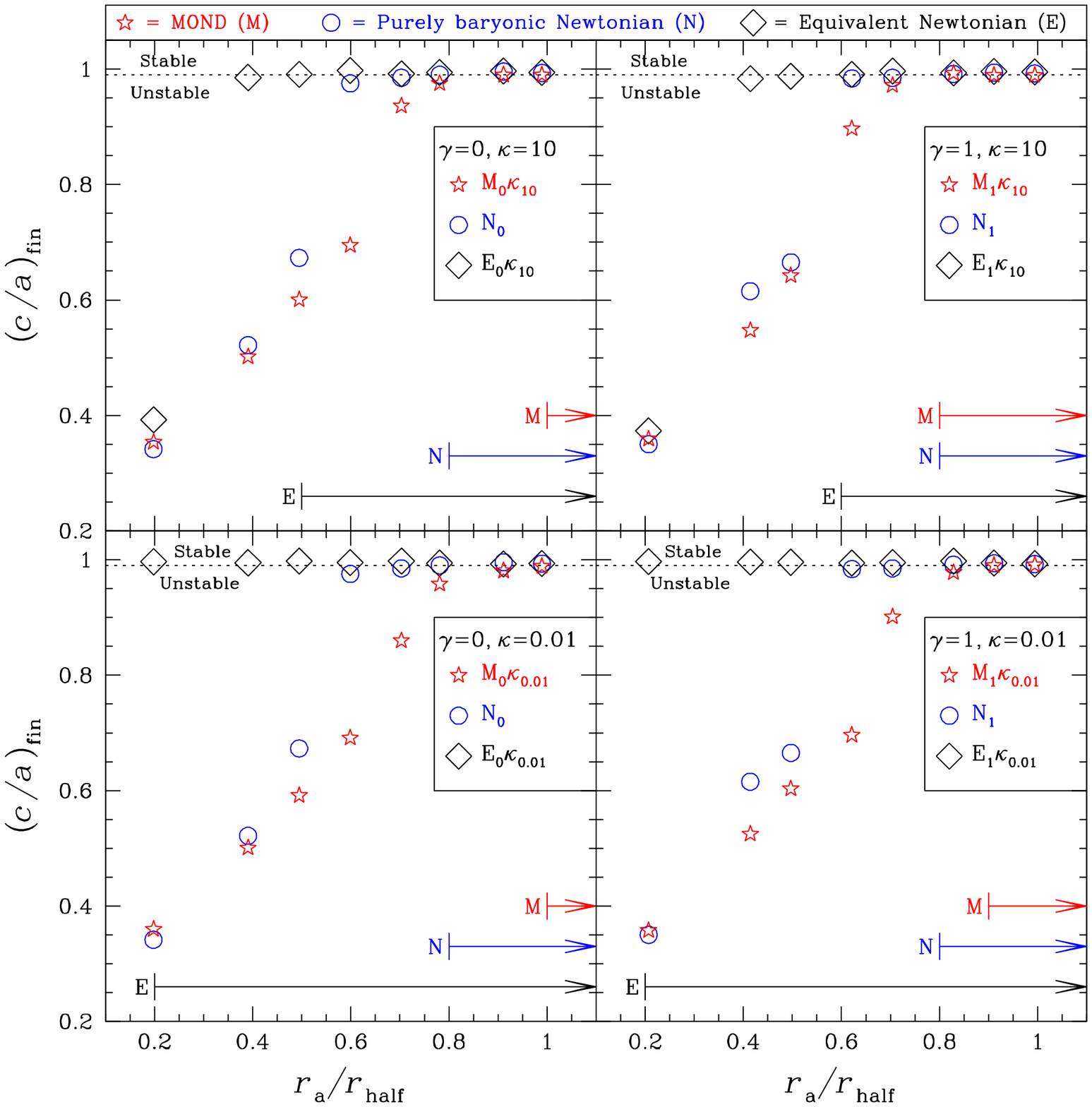,width=0.75\hsize}}
\caption{Final axis ratio $\cafin$ as a function of the initial
  anisotropy radius normalized to the half-mass radius for MOND
  (stars), purely baryonic Newtonian (circles), and ENSs (squares)
  with different values of $\gamma$ and $\kappa$. In each panel the
  horizontal dashed line marks the fiducial threshold value for
  stability, $\cafin=0.99$; the arrows indicate the range of
  $\ra/\rhalf$ corresponding to stable models.}
\label{figcara}
\end{figure*}

\section{Stability}
\label{secsta}

We now describe the main results of the $N$-body simulations. Before,
we briefly illustrate the numerical code \NMODY\ and the tests
performed.

\subsection{The numerical simulations}
\label{secsim}

We ran $N$-body simulations with \NMODY, our original parallel
three-dimensional particle-mesh code that can be used to follow the
evolution of either MOND or Newtonian collisionless stellar systems
\citep[][]{Nip07a,Lon09}.  In previous papers we have already used
\NMODY\ to demonstrate significant differences in violent relaxation,
merging and dynamical friction in MOND and Newtonian dynamics
\citep[][]{Nip07a,Nip07c,Cio07,Nip08}. We refer the readers to these
papers and to \citet{Lon09} for a more detailed description of the
code.  In the present study the spherical grid has 64 radial nodes, 32
nodes in colatitude $\vartheta$ and 64 nodes in azimuth $\varphi$, and
the total number of particles is $\Npart\simeq 8\times10^5$. We
verified with convergence experiments that these numbers of particles
and grid points exclude important discreteness effects.  In particular
we reran some of the simulations with a grid $128\times 64 \times 128$
and $\Npart\simeq 6.4\times10^6$, finding that the results are almost
identical to the corresponding lower-resolution cases. 

In each simulation the initial conditions consist of an $N$-body
realization of the stellar distribution of an isolated equilibrium
galaxy model, in which the particles are distributed in phase space
with the standard rejection technique, using the numerically recovered
distribution functions. In the ENS simulations, even if the DM halo is
frozen, the systems are truncated exponentially at a radius $\rt$, so
that the potential well is finite. As a rule, we adopt $\rt=10\rstar$,
but we ran some of the simulations also with $\rt=100\rstar$ and
$\rt=1000\rstar$, finding that the stability properties of a model do
not depend on the specific choice of the truncation radius.

Following \citet{Nip07c}, we identify each MOND initial condition by
fixing a value of the dimensionless internal acceleration parameter
$\kappa\equiv G\Mstar/\azero\rstar^2$, so $\Mstar$ and $\rstar$ are
not independent quantities. In physical units, introducing the
quantity $\Mstarten\equiv \Mstar/10^{10}\Msun$, $\rstar\simeq3.4
\kappa^{-1/2}\Mstarten^{1/2}\kpc$, the time and velocity units are
$\tstar\equiv\sqrt{\rstar^3/G\Mstar} \simeq
29.7\kappa^{-3/4}\Mstarten^{1/4}\Myr$, and $\vstar\equiv\rstar/\tstar
\simeq 112\kappa^{1/4}\Mstarten^{1/4}\kms$. The simulations are
evolved up to $100\tdyn$ with a time step $\Delta t =0.01 \tdyn$,
where $\tdyn$ is the characteristic dynamical time of the system.  In
the purely baryonic Newtonian $\gamma$-models, we adopt the standard
half-mass radius value
\begin{equation}
\tdyn \equiv \sqrt{\frac{3\pi}{16 G\rhohalf}}
=\frac{\pi\tstar}{\sqrt{2}\left[2^{1/(3-\gamma)}-1\right]^{3/2}},
\end{equation}
where $\rhohalf =3\Mstar /8\pi \rhalf^3$ is the mean stellar density
inside the half--mass radius $\rhalf$ (we recall that $\rhalf\simeq
3.84\rstar$ for $\gamma=0$ and $\rhalf\simeq 2.41\rstar$ for
$\gamma=1$).  In the corresponding MOND models (and their ENSs),
$\tdyn$ is given by the above expression multiplied by $\vcN/\vcMOND$,
where $\vcN$ is the circular velocity at $\rhalf$ for the purely
baryonic Newtonian system and $\vcMOND$ is the same quantity for the
MOND system.  

Following \citet{Nip02}, in order to determine whether a given model
is unstable, we check its departures from spherical symmetry by
monitoring the evolution of its intrinsic axis ratios $c/a$ and $b/a$
(where $a$, $b$ and $c$ are the longest, intermediate and shortest
axes of the inertia ellipsoid of the stellar distribution). As
preliminary tests we ran simulations in which the initial conditions
are isotropic purely baryonic Newtonian, MOND and ENSs. We found that
numerical uncertainties (due to the finite number of particles) on
$c/a$ and $b/a$ never exceed $1\%$ over $100\tdyn$ of evolution of
these stable isotropic models. As a consequence, we define {\it
  unstable} the models for which the axis ratio $\cafin$ after 100
$\tdyn$ is smaller than a fiducial threshold value $c/a=0.99$.  In the
simulations the onset of the instability is just due to numerical
noise produced by discreteness effects in the initial conditions.  As
expected, we found that an exact determination of the stability
threshold for a given family of models is not straightforward: while
for strongly anisotropic initial conditions the onset of the
instability is apparent and the numerical models settle down into a
final equilibrium configuration in a few dynamical times, for nearly
stable initial conditions the instability can be characterized by very
slow growth rates and its effects become evident even after tens of
$\tdyn$.

\begin{figure*}
\centerline{ \psfig{file=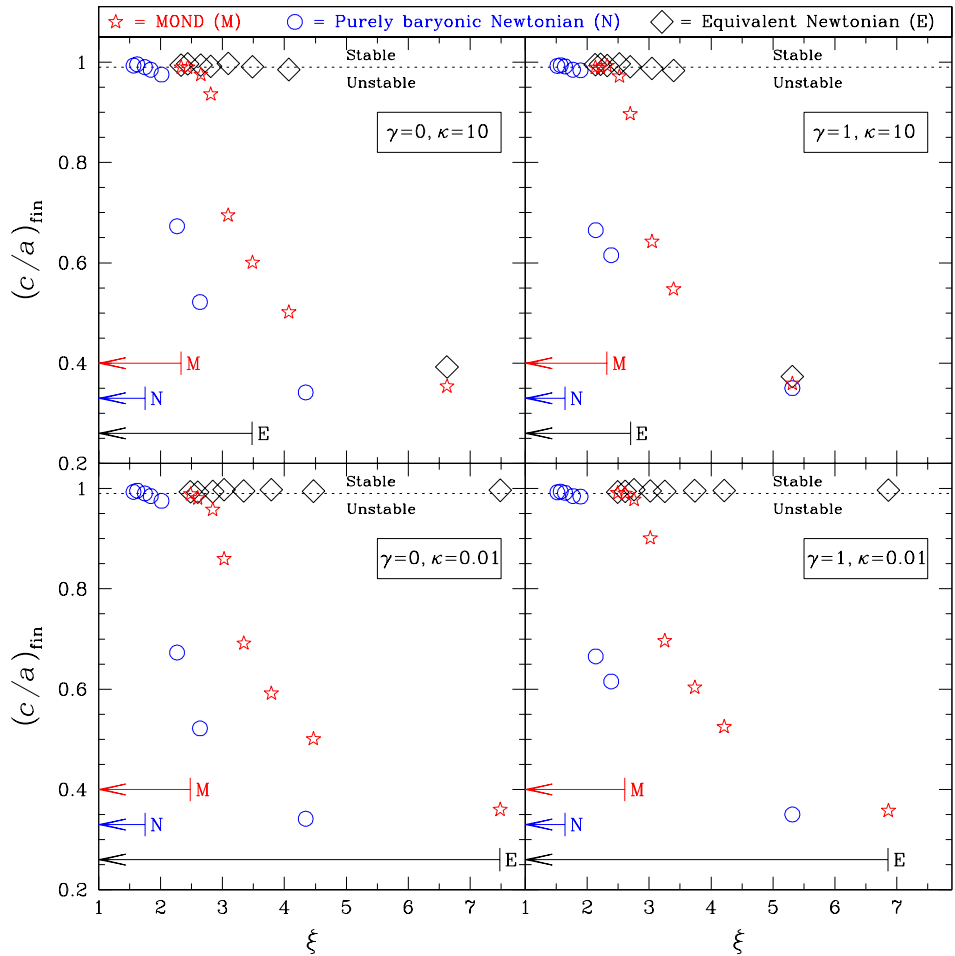,width=0.75\hsize}}
\caption{Final axis ratio $\cafin$ as a function of the
  Fridman-Polyachenko-Shukhman parameter $\xi$, for the same models as
  in Fig.~\ref{figcara}. The arrows indicate the range of $\xi$
  corresponding to stable models.}
\label{figxi}
\end{figure*}

\subsection{Results}
\label{secres}

For a given initial stellar density profile (i.e. $\gamma=0$ or
$\gamma=1$ in equation~\ref{eqrhostar}), we explored, besides the
purely baryonic Newtonian case (models N$_0$ and N$_1$ in Table~1),
two families of MOND systems (and the corresponding families of ENSs),
characterized by the values of the acceleration ratio $\kappa=0.01$
and $\kappa=10$ (see Table~1 for a summary).  The cases with
$\kappa=0.01$ correspond to low-acceleration systems with internal
accelerations everywhere much lower than $\azero$, in the so-called
deep-MOND regime. Their associated ENSs are therefore totally DM
dominated systems (Table~1, Column 3).  On the other hand, the
$\kappa=10$ cases are systems with relatively high accelerations
within $\rhalf$, corresponding in the Newtonian context to systems
dominated by baryons in the central regions and by DM only at $r\gsim
\rhalf$ (Table~1, Column 3). The four panels of Fig.~\ref{figden} show
the baryonic and DM initial density profiles of the ENSs corresponding
to the $\kappa=0.01$ and $\kappa=10$ MOND systems, for the two
explored values of $\gamma$. As can be seen, the associated DM halos
do not show a central hole and so, at least from this point of view,
are not unrealistic (see Section~\ref{seccon}).

\subsubsection{Minimum value for stability of the anisotropy radius}
\label{secra}

For each family of models we have a set of eight simulations, in which
the initial conditions differ only in the value of the normalized
anisotropy radius $\ra/\rhalf$. This is apparent in
Fig.~\ref{figcara}, where we plot the final axis ratio $\cafin$
vs. the initial value of $\ra/\rhalf$ for all the explored models of
the ten families.  The stability properties of the models can be
inferred directly from Fig.~\ref{figcara}: it is clear that the value
of $\ra$ at which instability appears is largest for MOND models,
smallest for the corresponding ENSs, and intermediate for purely
baryonic Newtonian systems.  Because of the very nature of the
  radial-orbit instability (very slow growth rates for marginally
  unstable systems), a precise determination of the value of the
  minimum anisotropy radius for stability $\ras$ is difficult. What
  can be estimated robustly with $N$-body simulations is a fiducial
  value of $\ras$ separating apparently unstable systems from {\it
    bona fide} stable systems. Adopting as fiducial threshold for
  stability $\cafin=0.99$ (horizontal dashed line in each panel of
  Fig.~\ref{figcara}), we estimate for each family of models the
  minimum anisotropy radius for stability $\ras$, which is reported
  (in units of $\rhalf$) in Table~1.

As expected, the differences in the stability limit between MOND and
ENSs are more evident in systems with lower values of the acceleration
ratio $\kappa$ (or, from a Newtonian point of view, for more DM
dominated systems).  For instance, if we consider $\gamma=0$ models
with $\kappa=0.01$, which are representative of low-surface density
systems with flat inner stellar density profile (such as dwarf
spheroidal galaxies), we find $\ras/\rhalf \sim 1$ for MOND and
$\ras/\rhalf<0.2$ for the ENS; very similar results are obtained for
the more peaked Hernquist $\gamma=1$ models. Unsurprisingly, Newtonian
models with the DM halo are less subject to the radial-orbit
instability than the purely baryonic Newtonian models.  

Thus, from this first analysis we conclude that, when using
$\ra/\rhalf$ as an indicator of the amount of admissible radial
orbits, MOND systems are more prone to radial-orbit instability than
their associated ENSs, and also than corresponding purely baryonic
Newtonian models.  However, as we will see in the next Section, the
comparison between the MOND and the associated purely baryonic
Newtonian families is subtle: the fact that the latter typically admit
smaller $\ra/\rhalf$ values does not imply that (globally) they can
sustain more radial kinetic energy.

\subsubsection{Maximum value for stability of the $\xi$ parameter}
\label{secxi}

Stability limitations expressed in terms of $\ra$ are particularly
relevant to observational works, as $\ra$ enters directly in the Jeans
equations that are routinely used to fit the velocity-dispersion
profiles of stellar systems. However, from a deeper point of view, the
value of $\ra$ (loosely speaking, the radius outside which orbits are
mainly radial) is not a robust measure of the fraction of kinetic
energy that is stored in radial orbits, which, at least in the
Newtonian context, is believed to be the main indicator of the
tendency to develop radial-orbit instability. More specifically, it
has been argued \citep[][]{Pol81,Fri84} that a proper criterion for
stability for Newtonian self-gravitating systems can be expressed in
terms of the global anisotropy parameter $\xi \equiv 2 \Tr/\Tt$, where
$\Tr$ and $\Tt \equiv \Ttheta+\Tphi$ are the radial and tangential
components of the kinetic energy tensor, respectively. Global isotropy
corresponds to a value of the Fridman-Polyachenko-Shukhman parameter
$\xi=1$. Indications exist that there is a critical value $\xis$ such
that only systems with $\xi\leq\xis$ are stable, and it is widely
accepted that $\xis \approx 1.7\pm0.25$, relatively independent of the
specific density distribution.

A priori we do not have reasons to expect that $\xi$ can be used as a
stability indicator also in MOND. In any case, by definition $\xis$
measures how much radial anisotropy can be supported by a stable
system, so it is interesting to discuss the stability properties of
our MOND, purely baryonic Newtonian and ENSs in terms of $\xi$.  The
values of $\xis$ for the families of models studied in the present
work are reported in Table~1, and in Fig.~\ref{figxi} we show all the
models in the $\cafin$-$\xi$ plane to be compared with the analogous
Fig.~\ref{figcara}.

For the purely baryonic Newtonian models the interpretation of these
numbers is straightforward: we find $\xis\sim1.6-1.8$, in agreement
with the standard criterion and with previous numerical studies
\citep[see, e.g.,][]{Nip02}.  For the MOND models we find $2.3\lsim
\xis\lsim2.6$, with $\xis$ depending on both the stellar density
profile and the acceleration ratio $\kappa$. These values are
substantially higher than those found for the corresponding purely
baryonic Newtonian models, indicating that, for given density
distribution, MOND systems can sustain more radial kinetic energy than
Newtonian systems without DM.  This is not in contrast with the
findings reported in the previous Section, i.e.  that MOND models have
larger $\ras$ than purely baryonic Newtonian models. In fact, for
fixed density profile and $\ra$ value, a MOND model has higher $\xi$
than a purely baryonic Newtonian model, because more kinetic energy is
stored in the outer parts, where orbits are radially biased and the
gravitational field of MOND systems is stronger\footnote{Note that by
  construction a MOND model and its ENS, with identical value of
  $\ra$, have the same value of $\xi$.}. Apparently, this effect
compensates the larger values of admissible $\ra$.  Instead, the ENSs
are again found able to sustain systematically higher values
of $\xis$ than the corresponding families MOND systems, and this
clearly indicates that MOND systems are more subject to radial-orbit
instability than Newtonian models with DM and identical total
gravitational field.

In general, for both MOND and ENSs we find a substantial spread in the
values of $\xis$, supporting the expectation that a ``universal''
value of $\xis$ does not exist for these systems \citep[for Newtonian
  models with DM, extended versions of the stability criterion have
  been proposed;][]{Pol87,Sti91}.  The data in Table~1 suggest that in
MOND, and even more in the ENSs, $\xis$ increases (i.e., relatively
more kinetic energy can be stored in radial orbits) for decreasing
$\kappa$. While for the ENSs this trend can be explained because in
the limit $\kappa\to 0$ the stars become just tracers, it is tempting
to speculate that in MOND this behaviour can be interpreted as a
manifestation of the less mixing nature of MOND with respect to
Newtonian gravity.  This interpretation is supported by the well
established numerical finding that phase mixing \citep{Cio07} and
violent relaxation \citep{Nip07a} in MOND systems take longer (in
units of dynamical times) than in the Newtonian case.  Very
qualitatively, the deep-MOND force between particles behaves like
$1/r$, i.e. it is nearer to the harmonic oscillator force than the
$1/r^2$ force, and it is easy to show that a system in which particles
interact with the harmonic oscillator force, no mixing or
instabilities are possible, as each particle moves independently of
the others \citep[][]{Lyn82}. Curiously, even though MOND forces in
a system of particles are non-additive, a similar trend towards longer
relaxation times has been also found recently in shell models
interacting with additive $1/r^{\alpha}$ forces \citep{DiC11}, and the
phase-space evolution of the case with $1/r$ forces is strikingly
similar to the deep-MOND collapses presented in \cite{Cio07}.

The fact that $\xi$ in MOND is strongly affected by the properties of
the system at large radii suggests to consider as a possible stability
indicator the quantity $\xihalf$, defined as the ratio of radial to
tangential kinetic energy within $\rhalf$: the maximum values for
stability $\xihalfs$ for our families of models are reported in the
last column of Table~1. Remarkably, all our MOND and purely baryonic
Newtonian families have $\xihalfs\sim1.3$, while the ENSs have
$\xihalfs\gsim1.5$ (for $\kappa=10$) and $\xihalfs\gsim3.6$ (for
$\kappa=0.01$), which leads us to speculate that {\it a limit on the
  amount of the radial orbits within the half-mass radius might be
  used as an empirical stability criterion for MOND stellar systems,
  valid from the Newtonian to the deep-MOND regime, i.e. independent
  of the value of the internal acceleration relative to $\azero$}.
This result is reminiscent of that of \citet[][]{Tre06}, who found
that an almost isotropic core can stabilize Newtonian self-gravitating
systems with very strong global radial anisotropy.

\begin{figure}
\centerline{ \psfig{file=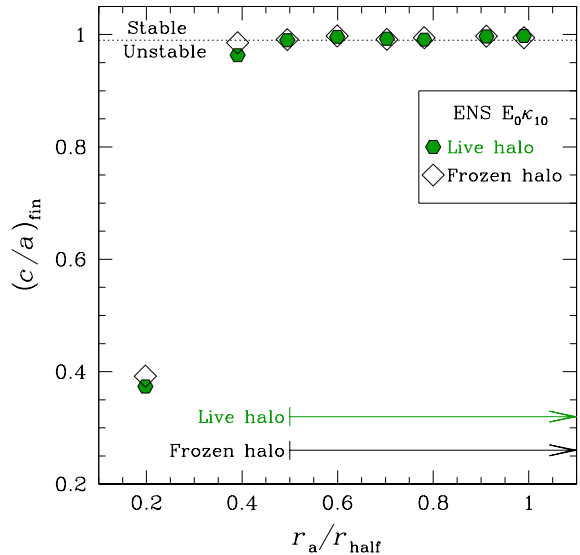,width=0.95\hsize}}
\caption{Final axis ratio $\cafin$ as a function of the initial
    anisotropy radius normalized to the half-mass radius for the ENSs
    with $\gamma=0$ and $\kappa=10$ when the DM halo is live
    (hexagons) and frozen (squares). The range of $\ra/\rhalf$
    corresponding to stable models (indicated by the arrows) is the
    same in the two cases.}
\label{figlive}
\end{figure}

\subsubsection{Simulations of equivalent Newtonian systems with live dark-matter halos}
\label{seclive}

The results on the stability of the ENSs might be affected to some
extent by our assumption of frozen DM halo, because it is possible
that an ENS with live DM halo has different stability properties
\citep[see][]{Sti91}.  In order to assess to what extent our results
are affected by the assumption of frozen DM halos, we reran the ENS
simulations of the family E$_0\kappa_{10}$ with live DM halos. For
technical reasons we ran these live-halo simulations with our
\FVFPS\ treecode \citep[Fortran Version of a Fast Poisson
  Solver;][]{Lon03,Nip03}, which was already tested against \NMODY\ in
\citet{Nip07c}. As an additional test, we reran with \FVFPS\ all the
purely baryonic Newtonian simulations of the family N$_0$, finding
excellent agreement with those run with \NMODY.  In the live-halo
simulations of the models of the family E$_0\kappa_{10}$ we used
$\simeq 8\times10^5$ particles for the stellar component and $\simeq
1.3\times10^6$ particles for the halo, which has mass $\simeq
1.6\Mstar$ for the adopted truncation radius $\rt=10\rstar$. We
verified numerically that for these two-component E$_0\kappa_{10}$
models the halo density distribution, which is monotonically
decreasing (see Fig.~\ref{figden}) and thus satisfies the necessary
condition for consistency (see Section~\ref{seccon}), can be generated
by a positive isotropic distribution function, which we used to
distribute the DM particles in phase space with the standard rejection
technique.

The final axis ratios of the live-halo simulations (hexagons in
Fig.~\ref{figlive}) are not significantly different from those of the
corresponding frozen-halo simulations (squares in Fig.~\ref{figlive}).
It follows that the stability properties of the E$_0\kappa_{10}$
family of models do not depend on whether the halo is modelled with
particles or it is a fixed potential: in particular the values of the
minimum anisotropy radius $\ras$ (and then of the maximum
Fridman-Polyachenko-Shukhman parameter $\xis$) for stability are the
same in the two cases.  These results suggest that our conclusion that
MOND models are substantially more subject to radial-orbit instability
than ENSs should be robust, at least if the DM halos are almost
isotropic.  Though the assumption of isotropy of the halos is not
necessarily justified, given that the anisotropy of DM distributions
is hard to constrain, it is clear that an isotropic halo can be always
be invoked, in the context of Newtonian gravity, to stabilize an
otherwise unstable strongly radially-anisotropic stellar system, while
this freedom is not allowed in the context of MOND.

\section{Discussion and conclusions}
\label{secfin}

To some extent, the MOND and DM interpretations of the kinematics of
galaxies can be considered degenerate, i.e.  several observational
features can be satisfactorily reproduced in both paradigms. This
raises interesting questions about possible tests to discriminate
between MOND and Newtonian gravity with DM.  Fortunately, it is now
well established that some important dynamical processes {\it are}
different, even in systems in which the total gravitational potentials
are identical. Examples are dynamical friction and two-body
relaxation.

In the present work we explored whether radial-orbit instability acts
differently in MOND and in Newtonian gravity (with and without DM); in
addition to the theoretical interest, this study can be useful to
constrain the interpretation of the observed kinematics of stellar
systems in the two cases. In particular, we have focused on the
stability of OM radially-anisotropic spherical $\gamma$-models
($\gamma=0$ and $\gamma=1$).  We compared the results obtained for
MOND models, ENSs and purely baryonic Newtonian systems, all with the
same stellar density distribution.  Overall, we found that MOND
systems are more prone to radial-orbit instability than their ENSs,
independent of the specific indicator ($\ra$ or $\xi$) used to
quantify the anisotropy. Compared to purely baryonic Newtonian systems
with the same density profile, however, MOND systems have larger
minimum anisotropy radius for stability $\ras$, but nevertheless
higher maximum global anisotropy parameter for stability $\xis$, a
consequence of the larger kinetic energy that can be stored in their
outer regions. We speculate that $\xis$ is larger in MOND systems than
in Newtonian systems with no DM for the same reasons that phase mixing
and violent relaxation are less efficient, i.e., because the force
between particles in deep-MOND decreases with distance less strongly
(qualitatively as $1/r$) than in Newtonian gravity, so being nearer to
the special case of harmonic oscillator inter-particle forces, when
instabilities and energy exchanges are impossible

Observationally, these findings may be relevant to applications of
MOND to pressure-supported systems such as globular clusters, dwarf
spheroidal and elliptical galaxies.  For instance, the
velocity-dispersion profiles of globular clusters in the outer parts
of the Milky Way can been used to test MOND. In the case of the
globular cluster NGC~2419 strong OM radial anisotropy ($\ra$ close to
$\rac$) might be required in order to bring the velocity dispersion
profile predicted by MOND close to the profile inferred from the
observed radial velocities of the cluster stars \citep{Sol10}.  A
combination of the stability constraints discussed in this paper with
new observations of the radial velocity of stars can make the
kinematics of NGC~2419 a crucial test for MOND (Ibata et al., in
preparation).

Another family of objects that are very interesting from the
perspective of MOND is that of dwarf spheroidal galaxies, because
their low surface-brightness, combined with their measured velocity
dispersion, lead to conclude that they must be DM dominated if
Newtonian gravity holds. \citet{Ang08} studied the observed
line-of-sight velocity-dispersion profiles of Milky Way dwarf
spheroidal galaxies, finding that in most cases the data can be
reproduced in MOND, at least with somewhat {\it ad hoc} anisotropy
profiles. This result should be reconsidered on the basis of the
consistency and stability constraints discussed in the present work.

The kinematics of elliptical galaxies has also been considered as a
possible test for MOND \citep[e.g.][]{San00}. As for globular clusters
and dwarf spheroidals, when trying to reproduce the kinematics of
ellipticals, the anisotropy of the velocity distribution of the stars
is one of the variables in the problem and it is important to
constrain it as much as possible, along the lines described in the
present paper.  For instance, consistency constraints (but not the
more stringent stability constraints) on the anisotropy of stars are
considered in the recent investigation by \citet{Car10}. Some authors
\citep{Tir07,Kly09} used the observed kinematics of planetary nebulae
or of the satellites around elliptical galaxies as a test for MOND,
allowing for quite general anisotropy profiles for the satellite
system. In this case, however, while some limits on the amount of
radial anisotropy might come from the requirement of consistency,
stability arguments do not necessarily lead to strong constraints as
long as the considered tracers are not dynamical dominant.

\section*{Acknowledgements}
Most numerical simulations were performed at CINECA, Bologna, with CPU
time assigned under the INAF-CINECA agreement 2008/2010.  L.C. and
C.N. are supported by the MIUR grant PRIN2008.

\end{document}